# Application of an Equilibrium Vaporization Model to the Ablation of Chondritic and Achondritic Meteoroids

Laura Schaefer, Bruce Fegley, Jr.
Planetary Chemistry Laboratory, McDonnell Center for the Space Sciences, Department of Earth and Planetary Sciences, Washington University, St. Louis, MO 63130-4899
laura_s@levee.wustl.edu, bfegley@levee.wustl.edu

**Abstract.** We modeled equilibrium vaporization of chondritic and achondritic materials using the MAGMA code. We calculated both instantaneous and integrated element abundances of Na, Mg, Ca, Al, Fe, Si, Ti, and K in chondritic and achondritic meteors. Our results are qualitatively consistent with observations of meteor spectra.

## Introduction

Identification of meteoroid compositions is hampered because most meteoroids do not reach the Earth's surface. We must therefore rely on observations of meteors, i.e. the hot gas ablated from a meteoroid, to tell us about the composition of meteoroids. However the composition of the meteor may not reflect the bulk composition of the meteoroid because of incomplete ablation. In this paper, we attempt to bridge the gap between the observed meteor spectra and the initial unobserved meteoroid composition by examining the vaporization chemistry of meteoroids.

Meteoroid ablation models usually focus on physical properties (initial mass, density, porosity) and motion (velocity, fragmentation) but do not consider the variable compositions of meteoroids (Pecina and Ceplecha 1983, Ceplecha et al. 1993, Zinn et al. 2004, Campbell-Brown & Koschny 2004). Instead a total vapor pressure is assumed for a generic "stony" object. Other models assume complete evaporation of CI chondritic material (Popova et al. 2001). However, observations show that meteors rarely have chondritic abundances of elements such as Ca, Al, and Ti (Ceplecha et al. 1998). Many meteor spectra are also time-variable, e.g., the abundance of Na in a meteor's spectra often decreases as the meteor descends through the Earth's atmosphere (Borovička et al. 1999).

McNeil et al. (1998) used the MAGMA code and developed a differential ablation model in which metals ablate sequentially along a meteor's trajectory based upon volatility. Their results help explain lidar observations of single- and double-element meteor trails of Na, K, Ca, $Ca^+$, and Fe, specifically the preponderance of single-element trails and the deviation of element ratios from CI chondritic values in the double-element trails (von Zahn et al. 2002; Murad & Roth, 2004).

In this paper, we use the MAGMA code to model fractional vaporization of chondritic and achondritic meteoroids. Our goal is to constrain the initial composition of meteoroids based upon observed elemental abundances in their meteor spectra. We explicitly assume that the vaporizing material does not interact with the atmosphere and that there are no kinetic deviations from equilibrium. We will model interaction with the atmosphere in later work. We also assume chemical equilibrium between the meteoroid and its evolved vapor. In later sections, we justify our assumptions and compare our results to observed meteor spectra. Our results complement those of McNeil and colleagues but do not duplicate their work.

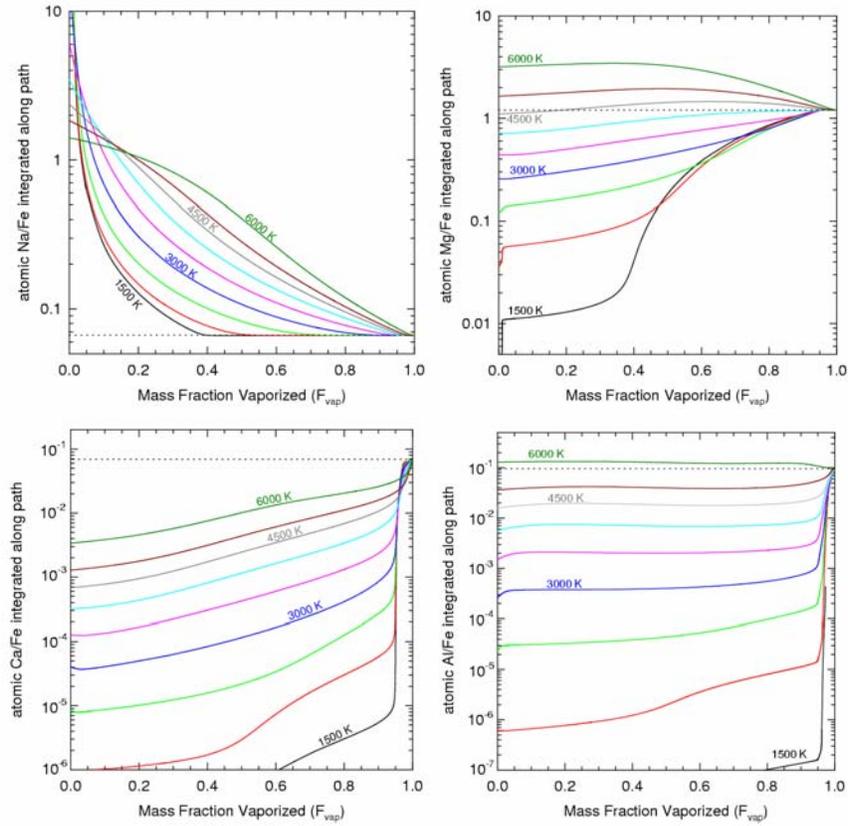

**Figure 1.** Integrated element abundances relative to Fe in the vapor over a CI chondritic melt, at various temperatures. Each line represents an isothermal vaporization path. In all figures the lines are spaced at 500 K intervals from 1500 K to 5000 K and the 6000 K isotherm is also shown. The dashed line marks the solar composition. (a) Na/Fe (b) Mg/Fe (c) Ca/Fe (d) Al/Fe.

## Methods

The MAGMA code is a mass balance, mass action code that computes fractional vaporization from a silicate melt composed of the oxides $SiO_2$, $MgO$, $FeO$, $Al_2O_3$, $CaO$, $TiO_2$, $Na_2O$, and $K_2O$. The MAGMA code uses the ideal mixing of complex components model developed in the 1980s by Hastie and colleagues at the U.S. National Bureau of Standards to simultaneously solve for the composition of the silicate melt, melt – vapor chemical equilibria, and vapor phase chemical equilibria. The code's operation, thermodynamic database, and validation against experimental studies are described in our prior papers and references cited therein (Fegley and Cameron 1987, Schaefer and Fegley 2004).

The MAGMA code calculates bulk element ratios in the vapor at individual points and integrated along a vaporization pathway, and gas speciation. We focus on the bulk element ratios in the vapor, both step-wise and integrated. Bulk element ratios refer to the ratio of the sums of the mole fractions for gas species of two elements; e.g. Mg/Fe refers to $\Sigma(X_{Mg} + X_{MgO}) / \Sigma(X_{Fe} + X_{FeO})$. Future work will describe gas speciation. The results are graphed as a function of the mass fraction of the melt that has been vaporized ($F_{vap}$). This quantity can be qualitatively identified with a decrease in altitude since a meteoroid loses mass as it descends through the atmosphere. A quantitative relationship between $F_{vap}$ and altitude, which depends upon such variables as meteoroid mass, meteor velocity, beginning height, etc., has previously been computed by Murad and colleagues (Murad & Roth 2004, McNeil et al. 1998). Such calculations are beyond the scope of this paper, so our comparisons to meteor spectra will necessarily be of a qualitative nature.

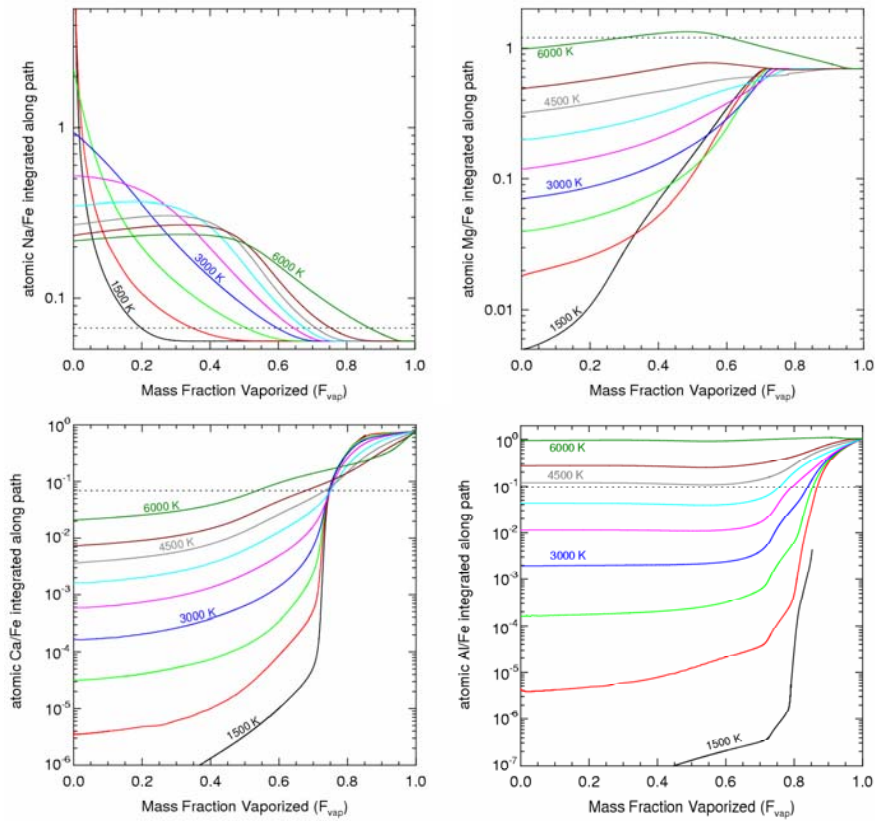

**Figure 2.** Integrated element abundances relative to Fe in the vapor over a eucritic melt, at various temperatures. (a) Na/Fe (b) Mg/Fe (c) Ca/Fe (d) Al/Fe.

The calculations shown here are for T = 1500 – 6000 K. All calculations are isothermal. We chose this temperature range based upon meteor observations. Ablation typically begins at temperatures of about 1500-2500 K (Ceplecha et al. 1998), and meteor spectra typically have temperatures in the range 4000-5000 K (Borovička 1993, 1994). Our temperature range spans these temperatures. Extreme high temperatures (> 4000 K) are shown for trends.

Meteoroids presumably have compositions similar to meteorites. The majority of meteorites (>70%) are chondrites, which are subdivided into ordinary, carbonaceous, and enstatite chondrites. All chondrites give very similar results; here, we show results for CI chondrites, which have solar abundances of the major non-volatile elements (Lodders 2003). Results for achondrites (eucrites, diogenites, aubrites) are more diverse. We show results for eucrites, which are the most abundant type of achondrite (~ 33% of all achondrites). We used the composition of the Juvinas eucrite, from Kitts and Lodders (1998). For both compositions, all iron is given as FeO (s), and the compositions have been normalized to 100 % on a volatile-free basis. Differences in results for chondrites and achondrites are discussed below.

## Results

Figures 1 and 2 show the bulk atomic ratios of Na, Mg, Ca, and Al relative to Fe integrated along the vaporization pathway above a CI chondritic meteoroid and a eucritic meteoroid, respectively. Figure 3 shows the bulk atomic ratios of Na and Ca relative to Fe at individual points along the vaporization pathway above a CI chondritic meteoroid. We chose Fe as the normalizing element because abundances determined from meteor spectra are typically given relative to Fe (e.g. Borovička 1993). We will discuss differences in results based upon the choice of element used for normalization. Below, we discuss the results for each element separately.

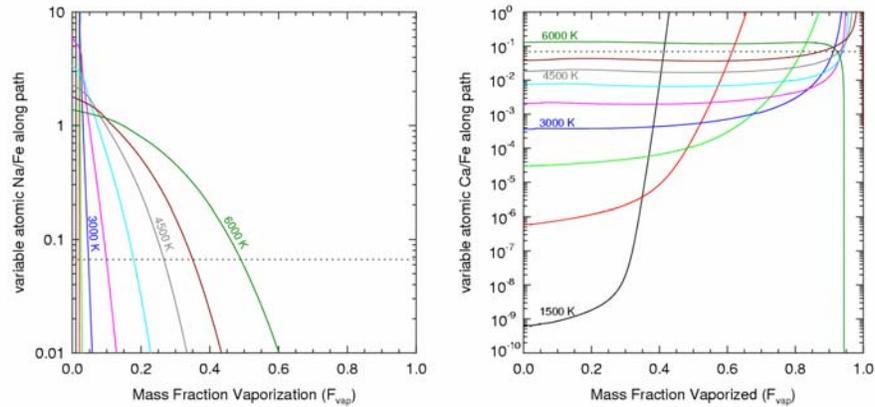

**Figure 3.** Instantaneous element abundances relative to Fe in the vapor over a CI chondritic melt, at various temperatures. (a) Na/Fe (b) Ca/Fe.

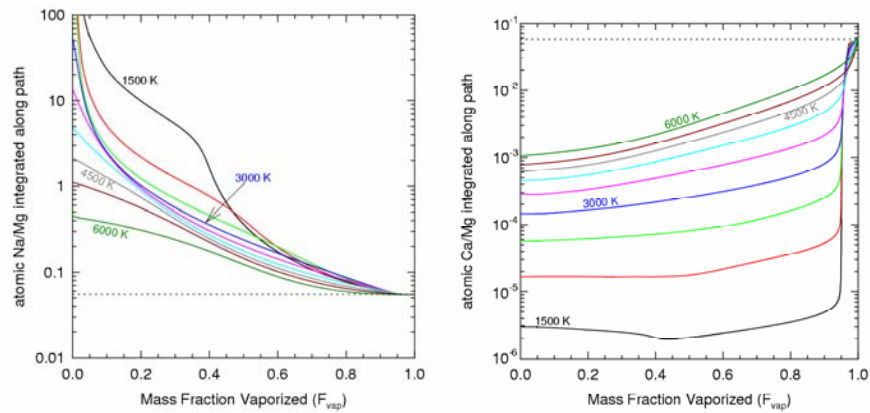

**Figure 4.** Integrated element abundances relative to Mg in the vapor over a CI chondritic melt, at various temperatures. (a) Na/Mg (b) Ca/Mg.

*Sodium.* In Fig. 1a and 2a the initial integrated atomic Na/Fe ratios at very low $F_{vap}$ are much larger than the solar value, especially at lower temperatures. As temperature increases, the initial Na/Fe ratio decreases to the solar value. For the CI chondrite, the Na/Fe ratio reaches solar values at ~40 wt% vaporized when T=1500 K; however, at 6000 K, the Na/Fe ratio does not reach the solar value until ~100% vaporization. For the eucrite, the atomic Na/Fe ratio drops below the solar value for $F_{vap}$ > 20% at 1500 K and > 90% at 6000 K. These trends indicate that the volatility of Fe increases as temperature increases. The atomic Na/Fe ratio over the eucrite is closer to the solar value than the CI chondrite value for most conditions (T and $F_{vap}$).

The initial step-wise atomic Na/Fe ratios shown in Fig. 3a are much larger than solar and rapidly decrease. The Na/Fe ratio drops more slowly as temperature increases for both chondritic and achondritic meteoroids. The curves for the atomic Na/Fe ratio above the eucrite (not shown) are much flatter than those of the CI chondrite, similar to the integrated ratios (Fig. 2a).

Figure 4 shows the integrated atomic Na/Mg and Ca/Mg ratios over a CI chondrite. For a given $F_{vap}$, the integrated Na/Fe atomic ratio over a chondrite increases as temperature increases (Fig. 1a), whereas the integrated Na/Mg atomic ratio decreases (Fig. 4a). At a given temperature, the Na/Mg ratio approaches solar values at much higher $F_{vap}$ compared to the Na/Fe ratio. The difference between ratios relative to Fe and Mg is most extreme for volatile elements (e.g., Na, K), and less important for refractory elements (e.g., Ca, Al). The integrated Ca/Mg atomic ratios (Fig. 4b) increase slightly less rapidly than Ca/Fe as temperature increases.

*Magnesium.* In Fig. 1b and 2b the integrated atomic Mg/Fe ratio generally increases as temperature and $F_{vap}$ increase. Magnesium is depleted over the eucrite compared to the CI chondrite. The atomic Mg/Fe ratio over the CI chondrite is about solar for T = 4000 – 4500 K. At T>4500 K, Mg/Fe is greater than solar values, and at T<4000 K, Mg/Fe is less than solar. The atomic Mg/Fe ratio over the eucrite is solar at ~6000 K. For T < 6000 K, the Mg/Fe ratio over the eucrite is subsolar.

Though not shown, we also calculated the step-wise atomic Mg/Fe ratios for both the CI chondrite and the eucrite. Generally, the Mg/Fe ratio increases with $F_{vap}$ because Mg is less volatile than Fe. At low temperatures, the Mg/Fe ratio starts out at ~1% of solar and quickly skyrockets to values greater than 1000 as all of the Fe is vaporized. As temperature increases, the curves begin to level out, similar to the Ca/Fe ratios shown in Fig. 3b. Between 4000 – 4500 K, the Mg/Fe ratio over the CI chondrite is about solar for all $F_{vap}$ < 0.8. The Mg/Fe ratio above the eucrite shows similar behavior but is shifted down because the eucrite is Mg-poor relative to the CI chondrite. The eucrite has solar Mg/Fe ratios at 6000 K for all $F_{vap}$ < 0.5.

*Calcium.* In Fig. 1c and 2c the integrated atomic Ca/Fe ratios increase with temperature and $F_{vap}$. The eucrite is a Ca-rich achondrite, with more Ca than the CI chondrite, so the Ca/Fe ratios above the eucrite are larger than those above the CI chondrite. The atomic Ca/Fe ratio over the chondrite approaches the solar value for $F_{vap}$ > 0.90. The atomic Ca/Fe ratio over the eucrite is approximately solar at T > 6000 K and exceeds the solar value for all $F_{vap}$ > 0.70.

In Fig. 3b the initial step-wise atomic Ca/Fe ratios over a CI chondrite are much less than solar at low temperatures and increase with $F_{vap}$. The abrupt increase in the Ca/Fe ratio at low temperatures is due to the early loss of all Fe from the melt. At higher temperatures, the stepwise Ca/Fe ratio flattens, and is fairly constant for most $F_{vap}$. It has an approximately solar value between 5000 and 6000 K. Similar behavior is seen above a eucritic meteoroid, although the Ca/Fe ratio is significantly higher, due to the larger initial Ca content of the eucrite. The Ca/Fe ratio above the eucritic meteoroid is approximately solar between 4000 and 4500 K.

*Aluminum.* As Fig. 1d and 2d show, there is essentially no Al in the vapor at low temperatures. The atomic Al/Fe ratio increases with temperature and $F_{vap}$. The eucrite is Al-rich compared to the CI chondrite. Therefore, the Al/Fe ratios above the eucrite are larger than those above the CI chondrite. The atomic Al/Fe ratio over the CI chondrite is approximately solar between 5000 and 6000 K. The atomic Al/Fe ratio over the eucrite is approximately solar between 4000 and 4500 K.

As with Mg and Ca, the step-wise atomic Al/Fe ratios begin at values much lower than solar and increase with temperature and $F_{vap}$. As temperature increases, the Al/Fe lines flatten, but they still increase slightly with $F_{vap}$ unlike the Ca/Fe ratios in Fig. 3b. The Al/Fe ratios over the CI chondrite only approach the solar value at $F_{vap} > 0.90$. At 6000 K, the Al/Fe ratio over the CI chondrite is $3 - 10\%$ of the solar value for all $F_{vap} < 0.90$. The Al/Fe ratio over the eucrite is slightly higher at all temperatures and is greater than the solar value for all $F_{vap} > 0.5$.

**Discussion**

*Equilibrium vs. Kinetics* Modeling the ablation of a meteoroid has two extreme end members: (a) ablation controlled entirely by kinetics (e.g., sputtering, chemical kinetics) or (b) ablation controlled entirely by equilibrium chemistry (vaporization). Most ablation models are of the first type, focusing purely on kinetics. Our model assumes that equilibrium chemistry is all-important. True meteor ablation is certain to be a combination of both kinetic and equilibrium effects. However, in order to identify deviations from equilibrium, we must first study pure equilibrium. By knowing the equilibrium composition, we may be able to determine the extent to which chemical and physical kinetic effects cause deviations from equilibrium.

We looked at chemical lifetimes for the vaporized species to see if chemical equilibrium is attained within typical meteor flight times. For example:

$$Si + O_2 = SiO + O \qquad (1)$$

has a chemical rate constant (Le Picard et al., 2001) of:

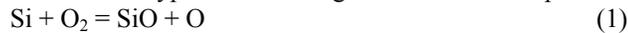
$$k_1 = 1.72 \times 10^{-10} (T/300 \text{ K})^{-0.53} \exp(-17/T) \text{ cm}^3 \text{ s}^{-1}. \qquad (2)$$

Using our calculated $O_2$ partial pressure from vaporization, we calculate a chemical lifetime for Si (g) of ~0.008 s at 1500 K, ~$10^{-9}$ s at 4000 K, and ~$10^{-11}$ s at 6000 K. These lifetimes are much shorter than the typical flight time of a meteor (~0.1-0.5 s), so chemical equilibrium may occur in the gas surrounding the meteoroid. We also assume chemical equilibrium within the meteoroid itself, which is valid for meteoroids that are small enough to melt completely (0.05-0.5 mm radius, Ceplecha et al. 1998) because diffusion and heat conduction do not significantly effect vaporization. Most micrometeoroids <100 μm in diameter will not reach temperatures above ~2000 K, and they may only experience partial melting or no melting at all, so equilibrium vaporization may not be valid for them. Micrometeoroids between 100 - 1000 μm in diameter can reach temperatures up to ~3000 K and many completely evaporate (Love & Brownlee 1991). Centimeter and greater -sized meteoroids may reach higher temperatures due to compressional shock heating, but do not melt completely; in these meteoroids, melt layers are typically confined to "skins" a few millimeters thick (Ceplecha et al. 1998). Diffusion and heat conduction are important in these larger meteoroids; therefore, chemical equilibrium is a less valid assumption.

*Comparison of results to observations of meteor spectra* Our model is generally consistent with observations of the elemental abundances in Leonid meteors (Borovička et al. 1999), in which Na peaks at the beginning of the meteor's flight path and rapidly decreases, while Mg peaks later in the flight. Borovička et al. (1999) show the time-variable abundances of Na and Mg in a variety of other meteoroids in their Fig. 6. Some of these meteors are consistent with our calculated curves for a CI chondrite, whereas others, such as one of the Taurids (SZ 28) and one of the sporadics (SZ 132), are more consistent with our calculated curves for a eucrite. This is primarily due to the high and relatively constant Na abundances in the meteors. As discussed above, the eucrite meteoroid has fairly constant Na/Fe ratios, which qualitatively agrees with the Na/Fe ratios observed in SZ 28 and SZ 132. The difference in the Na/Fe ratios between the CI chondrite and the eucrite is primarily due to the eucrite's lower initial Na abundance. Therefore, we suggest that these meteors had an initial Na abundance less than that of our CI chondrite.

Trigo-Rodriguez et al. (2004) show time-variable elemental abundances for a Leonid meteor (LEO) in their Fig.4. Their spectra show that the Na/Si ratio is fairly constant over most of the meteor's height for temperatures of 4400 K to 5000 K, and then it decreases as temperatures reach 5800 K. According to our results, this is consistent with an initial meteoroid slightly depleted in Na relative to our CI chondrite. Trigo-Rodriguez et al. (2004) suggest that the meteors they studied are enriched in Na relative to solar; we suggest that this may not be true unless there is an accompanying enrichment in Si. Figure 4 of Trigo-Rodriguez et al. (2004) also shows the time-variable Mg/Si, Ca/Si, and Fe/Si ratios for the same meteor (LEO). Comparing these abundances to our results, we find that the meteoroid may have larger Mg, Ca and Fe abundances and/or a smaller Si abundance relative to our CI chondrite in order to get similar elemental ratios in the gas.

Some meteors show complex behavior not explained by our model. These spectra can often be explained if the initial spectra is due to ablation of a fusion crust (for larger meteors), and spikes in the element abundances are due to sudden fragmentation of the meteoroid and ablation of material that was previously protected from ablation by the fusion crust. For example, Borovička (1993) analyzed the spectrum of a fireball, which is a meteor of large magnitude associated with a large meteoroid. In Fig. 14 of this paper, Borovička gives the time-variable element to Fe ratios as a function of height. The abundances (e.g., Ca, Ti, Cr, Na) show several peaks, which indicate a sudden increase in vaporization and which are accompanied by an increase in the meteor's temperature. Our results cannot reproduce such a spike in the element abundances unless we overlay a separate vaporization path for a body of the original composition at a higher temperature, which vaporizes at a much faster rate. This is consistent with Borovička's conclusion that the flares correspond to fragmentation events.

A caveat about interpretation: Here we present bulk elemental abundances in the vaporized gas, i.e., the abundance of Na includes the species Na, $Na_2$, NaO, $Na_2O$, and $Na^+$. Observations, however, typically only give abundances for the neutral (and possibly singly ionized) monatomic gases. The outcome is that the results presented in our figures may differ slightly from observed meteor abundances for a meteoroid of the same composition. The discrepancy depends strongly on the element and is most severe for Al, which occurs in the gas primarily as AlO (g). The gas speciation is computed by the MAGMA code and will be given elsewhere.

**Summary**

We applied results from the MAGMA code, an equilibrium vaporization model, to the problem of meteoroid ablation. In particular, we looked at the element ratios in the meteor. Hopefully, this data can be used to estimate the original meteoroid composition from meteor observations. In future work, we will compare the calculated melt compositions from MAGMA to analyzed compositions of micrometeorites. In principle, this will tell us if equilibrium is a valid assumption when studying meteoroid vaporization. We will also add additional compounds and elements of interest, such as $H_2O$, S, C, Mn, Co, Cr, Ni, etc., which are observed in meteor spectra, and other trace elements that are analyzed in micrometeorites. In this way, we may be able to make greater distinctions between types of meteoroids based upon meteor spectra. Additionally, we plan to model interaction of the vaporized gas with hot air (e.g., $N_2$, $O_2$, $CO_2$, etc.), which will be useful for studying meteor trains. The MAGMA code is freely available upon request.

**Acknowledgements**

This work is supported by the McDonnell Center for the Space Sciences and Grant NNG04G157A from the NASA Astrobiology Institute.